\documentclass[final,1p,times]{elsarticle}

\usepackage{amssymb}
\usepackage{amsmath}
\DeclareMathOperator{\sech}{sech}

\usepackage[dvipsnames]{xcolor}
\definecolor{bg}{RGB}{239, 241, 241}
\definecolor{grn}{rgb}{0,0.6,0}
\definecolor{mve}{rgb}{0.58,0,0.82}

\usepackage{listings}
\usepackage{url}

\journal{Journal of Computational Physics}

\begin{document}

\begin{frontmatter}



\title{High-Performance Computational Magnetohydrodynamics with Python} 


\author[1]{C. Bard} 
\author[1]{J. Dorelli}

\affiliation[1]{organization={Heliophysics Division, NASA Goddard Space Flight Center},
            addressline={},
            city={Greenbelt},
            postcode={},
            state={Maryland},
            country={USA}}

\begin{abstract}
We present the AGATE simulation code, a Python-based framework developed primarily for solving the magnetohydrodynamics (MHD) equations while maintaining adaptability to other equation sets.
The code employs a modular, object-oriented architecture that separates interface specifications from numerical implementations, allowing users to customize numerical methods and physics models.
Built on a Godunov-type finite-volume scheme, AGATE currently supports the ideal, Hall, and Chew-Goldberger-Low (CGL) MHD equations, with multiple acceleration options ranging from Numpy to GPU-enabled computation via NVIDIA CUDA.
Performance testing demonstrates that our GPU implementations achieve $>$40x speedups over CPU versions.
Comprehensive validation through established benchmarks confirms accurate reproduction of both linear and nonlinear phenomena across different MHD regimes.
This combination of modularity, performance, and extensibility makes AGATE suitable for multiple applications: from rapid prototyping to production simulations, and from numerical algorithm development to physics education.
\end{abstract}

\begin{graphicalabstract}
\end{graphicalabstract}

\begin{highlights}
\item AGATE introduces a modular Python framework for MHD simulations
\item Modular design allows straightforward implementation of new numerical methods and physics models
\item Multiple acceleration options support both CPU and GPU computation, with GPU implementations achieving >40x speedups over CPU versions
\end{highlights}


\end{frontmatter}



\section{Introduction}
\label{sec:Intro}

High-performance computing (HPC) in physics increasingly demands substantial computational resources, with recent models reaching exascale capabilities \cite{exascale18K,exascale22E,exascale22J,exascale2023C,dongarra2024co}.
These advanced simulations typically rely on low-level programming languages for optimal performance.
In computational physics, codes such as Kokkos \cite{kokkos22}, Parthenon \cite{grete2023parthenon}, FLASH \cite{FLASH14D,FLASH22D}, and AMREX \cite{amrex21Z} utilize C++ or Fortran.
Similarly, heliophysics codes such as GAMERA \cite{zhang2019gamera}, Space Weather Modeling Framework \cite{SWMF05}, and OpenGGCM \cite{openggcm08R} employ Fortran, while COCONUT \cite{coconut22P} and EUHFORIA \cite{pomoell2018euhforia} use C++.

While these low-level languages enable high performance, they present significant barriers to the average scientist.
As an alternative, Python has emerged as one of today's most popular programming languages \cite{TIOBE}, offering easier development and debugging compared to C or Fortran.
Recognizing this, some HPC frameworks offer Python interfaces (e.g., AMREX/PyAMREX \cite{amrex24python}); however, modifying or extending the underlying code remains challenging.
Furthermore, Python's traditionally slower execution speed has limited its adoption in HPC calculations.

Recent efforts have produced several solutions to accelerate Python performance: Numpy \cite{numpy20} utilizes vectorized array operations, Numba \cite{numba15} generates machine code to use multi-threading, and Cupy \cite{cupy17} provides GPU acceleration through NVIDIA CUDA libraries.

Leveraging these modern Python acceleration technologies, we present AGATE, a new HPC simulation code written entirely in Python.
As a first step, we focus on a finite-volume scheme solving the ideal, Hall, and Chew-Goldberger-Low (CGL) magnetohydrodynamics equations.
The framework employs a modular design that enables users to implement custom equations and methods, or use default settings.
The use of Python also makes AGATE well-suited for diverse applications beyond HPC, serving as an effective platform for numerical algorithm development, physics education, and rapid prototyping.

The paper is organized as follows: Section \ref{sec:modular} describes AGATE's modular organization and implementation of a Godunov-type finite-volume scheme; Section \ref{sec:MHD} presents the mathematical formulation for the ideal, Hall, and CGL MHD equations; and Section \ref{sec:benchs} validates the implementation through standard benchmarks and analyzes performance across different acceleration options.

\section{The AGATE Code}
\label{sec:modular}
\subsection{Modular Organization}
Python's object-oriented structure enables the creation of user-defined types through classes.
AGATE employs this design philosophy by structuring its numerical solver as a collection of distinct Python classes.
The framework uses a dual-class architecture to encapsulate each step as an Interface paired with an Implementation.
Interfaces manage high-level program operations, shielding the main program from the underlying numerical complexities, while Implementations handle the detailed calculations.
Users can select or create desired Implementations; the solver will operate correctly as long as these implementations adhere to their corresponding Interface specifications.

\begin{figure}[t]
\centering
\includegraphics[width=0.9\textwidth]{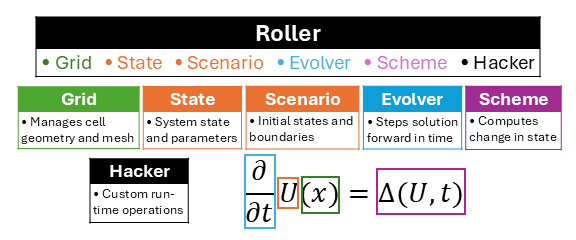}
\caption{AGATE's top-level class hierarchy.
Classes highlighted in color correspond to the components of the numerical equations highlighted by the same color.
Detailed descriptions of each class are provided in Section \ref{sec:modular}.}\label{fig:AGATEclasses}
\end{figure}

In general, AGATE solves the equation $\partial U / \partial t = \Delta$ with $\Delta = \Delta(U,t)$ the instantaneous rate of change of the state variable vector $U$ at time $t$.
This equation is discretized on a numerical grid, managed by the \texttt{Grid} class, which stores key information regarding cell locations, sizes, and connectivities.
At the moment, only an uniform Cartesian grid is supported; however, AGATE's modular design allows for straightforward implementation of additional grid structures.

The \texttt{Vector} class handles the state variable $U$ by storing values at corresponding \texttt{Grid} locations.
For MHD simulations, the \texttt{State} class extends \texttt{Vector} with enhanced capabilities, including pointers to specific MHD variables and storage for non-state variables such as the ratio of specific heats $\gamma$ for isothermal MHD.

The \texttt{Evolver} class manages the time evolution operator $\partial/\partial t$.
Our current implementation offers several Runge-Kutta schemes (e.g. \cite{butcher96history, butcher16text}), ranging from RK11 (1-stage, first-order) and RK22 (2-stage, second-order) to more sophisticated strong-stability preserving (SSP) methods. These SSP methods include RK33 \cite{shu88, gottlieb98}, as well as higher-order RK43 and RK53 schemes \cite{macdonald03}.

Additionally, \texttt{Evolver} manages the calculation of $\Delta$ via the \texttt{Scheme} interface class.
The framework currently implements three magnetohydrodynamics equation sets: ideal MHD, Hall MHD, and the Chew-Goldberger-Levy (CGL) anisotropic MHD equations \cite{chew1956}.
More details of the MHD implementations are provided in Section \ref{sec:MHD}.

AGATE also includes utility classes to streamline common numerical simulation tasks.
The \texttt{Scenario} class handles initialization by defining problem-specific initial conditions and grid boundaries, while providing methods to create properly initialized \texttt{State} instances associated with given \texttt{Grid} objects.
The \texttt{Hacker} class enables users to implement custom, condition-triggered tasks during simulation execution, supporting capabilities from real-time analysis to runtime modifications.
For instance, in the GEM benchmark discussed below (Section \ref{sec:GEMbench}), we wrote a cubstom \texttt{Hacker} class to calculate reconnection rates at specified intervals during runtime, eliminating the need for post-processing of saved output files.

Finally, the \texttt{Roller} class serves as AGATE's top-level controller, managing the time evolution loop that advances the state variable $U$ from its initial configuration $U_0$ at time $t_0$ to its final state $U_1$ at time $t_1$.
While users can manually configure individual algorithm components, \texttt{Roller} provides "autoinit" functions that streamline the setup process by automatically initializing all necessary algorithm classes using a combination of preset defaults and user-specified options.
This approach offers flexibility, allowing users to choose between convenient automated initialization and detailed manual control of the numerical algorithm's components.

These classes are summarized in Figure \ref{fig:AGATEclasses}.

\subsection{Implementation of Finite-Volume Scheme}
As mentioned above, AGATE encapsulates the numerical scheme's core calculations within the \texttt{Scheme} class.
The default implementation solves $\partial U / \partial t = \nabla\cdot \vec{F}(U)$ using a general Godunov type finite-volume (FV) scheme \cite{godunov59}, discretized via:
\begin{align}
U^{n+1}_{ijk} = U^n_{ijk} - \frac{\Delta t}{\Delta x}\left(F^n_{i+1/2,j,k} - F^n_{i-1/2,j,k}\right) - \frac{\Delta t}{\Delta y}\left(G^n_{i,j+1/2,k} - G^n_{i,j-1/2,k}\right) - \frac{\Delta t}{\Delta z}\left(H^n_{i,j,k+1/2} - H^n_{i,j,k-1/2}\right)
\end{align}
$U^n_{i,j,k}$ is the solution state in cell $(i,j,k)$ at timestep $n$ with numerical flux vector components $F,G,H$ calculated at their respective cell faces $i,j,k~\pm1/2$.
The spatial discretization uses cell lengths $\Delta x$, $\Delta y$, $\Delta z$, while the timestep $\Delta t$ is determined by the global Courant condition: $\Delta t = C\Delta x_{min}/v_{max}$.
$\Delta x_{min}$ is the smallest cell length in the simulation grid over all dimensions and $v_{max}$ is the fastest problem-specific wavespeed in the domain.
The Courant parameter $C$ defaults to dimension-specific values (0.8 for 1D, 0.45 for 2D, and 0.325 for 3D), though other values can be set.

The FV scheme incorporates distinct subclasses that are each Interfaces for a step in the solution process:
\begin{itemize}
    \item \texttt{Reconstructor}; handles reconstruction on cell interfaces
    \item \texttt{Speeder}: calculates wave speeds
    \item \texttt{Fluxor}: evaluates physical fluxes
    \item \texttt{Riemann}: solves the Riemann problem
    \item \texttt{Sourceror}: calculates source terms
\end{itemize}
Each Interface couples with corresponding Implementations to provide several options for calculation.
For example, the \texttt{Reconstructor} may use a first-order Godunov scheme or a higher-order reconstruction.

In the following paragraphs, we delineate the default choices for the FV scheme, though we mention that AGATE's modular organization makes it easy to substitute other options.

First, we implement a second order scheme for reconstructing the cell interface state $U^L$ and $U^R$:
\begin{align}
U^{L}_{i+1/2} = U_i + \Delta U_i/2\\
U^{R}_{i-1/2} = U_i - \Delta U_i/2
\end{align}
Here, superscripts $L,R$ denote interface values from the (L)eft or (R)ight of interface $i\pm1/2$, noting that interface $i+1/2$ is the right face of cell center $i$.
Similar equations can be defined for the $y,z$ faces in indices $j$ and $k$.
The slopes $\Delta U$ are calculated using the monotonized central (MC) limiter (e.g. \cite{toth08})
\begin{align}
\Delta U_i = \mathrm{minmod}\left[\beta(U_{i+1} - U_i), \beta(U_{i} - U_{i-1}), \frac{U_{i+1}-U_{i-1}}{2}\right]
\end{align}
with $\beta$ defaulting to a value of 1.3.
The minmod function returns zero if the arguments have different signs, otherwise it returns the argument with the smallest magnitude.

The numerical fluxes $F,G,H$ are calculated using the HLL approximate Riemann solver \cite{harten83, toro99}.
At interface $i+1/2$, the flux is given by:
\begin{align}
F_{i+1/2} = \frac{s'_R \phi_L - s'_L \phi_R + s'_L s'_R (U^R - U^L)}{s'_R - s'_L}
\end{align}
with similar equations for $G$,$H$.
$\phi_{L,R}$ are the physical fluxes calculated directly from the interface states $U^L~,U^R$ using the appropriate set of equations, such as MHD (equations \ref{eq:MHDeqns} - \ref{eq:divclean}).

The modified wavespeeds $s'_L = \min(s_L, 0)$ and $s'_R = \max(s_R,0)$ derive from the interface signal speeds $s_{L,R}$, e.g., \cite{davis88}:
\begin{align}
s_L = \min(v_L - c_L, v_R - c_R)~;~s_R = \max(v_L + c_L, v_R + c_R)
\end{align}
Here, $c_L$ and $c_R$ are problem-specific wavespeeds calculated from $U_L$ and $U_R$.
Sections \ref{sec:idMHD} and \ref{sec:CGLMHD} provide comprehensive descriptions of these wavespeeds, along with the fluxes and source terms for ideal, Hall, and CGL MHD.

\subsection{Modes of Operation}
\label{sec:acclr}
AGATE offers multiple acceleration options to optimize performance (or convenience) across different computing environments: \texttt{None}, \texttt{Numpy}, \texttt{Numba}, \texttt{Cupy}, and \texttt{Kernel}.
These options determine which libraries handle the low-level numerical calculations within the simulation.
The first three options (\texttt{None}, \texttt{Numpy}, \texttt{Numba}) operate on CPU architectures, while \texttt{Cupy}, and \texttt{Kernel} leverage GPU acceleration via the CuPy Python interface to the NVIDIA CUDA library.
AGATE automatically configures the appropriate computational functions during initialization.
All other code dealing with top-level algorithms and program organization (Section \ref{sec:modular}) is identical across all acceleration modes.
For more details and examples concerning the low-level differences between libraries, see \ref{app:acclr}.

\section{Solving the MHD Equations}
\label{sec:MHD}
\subsection{Ideal and Hall MHD}
\label{sec:idMHD}
AGATE provides a solver for the magnetohydrodynamics (MHD) equations in the ideal, Hall, and CGL limits.
In these equations, the density ($\rho$), magnetic field ($\mathbf{B}$), and length scale are normalized using reference values $\rho_0$, $B_0$, and $L_0$, respectively.
The bulk velocity $\mathbf{v}$ is normalized to the Alfv\'en speed: $v_0 = v_{A0} = B_0/\sqrt{4\pi\rho_0}$.
The pressure is normalized with $P_0 = B_0^2/(4\pi)$, and time is normalized via $t_0 = L_0/v_0$.
Additionally, since the default MHD equations do not necessarily handle $\nabla\cdot\mathbf{B}$ divergence errors (e.g. nonphysical acceleration along field lines \cite{brackbill80}), AGATE also provides classes to solve the Generalized Lagrangian Multiplier (GLM) formulation of MHD \cite{dedner02}.
GLM-MHD provides a mechanism for "cleaning" the magnetic divergence via damping and propagating the errors out of the domain via a scalar function $\psi$ whose evolution is constructed to be identical to $\nabla\cdot\mathbf{B}$.

For ideal and Hall MHD, the resulting equations are:
\begin{align}
\label{eq:MHDeqns}
\frac{\partial\rho}{\partial t}& + \nabla\cdot(\rho \textbf{v}) = 0,\\
\frac{\partial\rho\textbf{v}}{\partial t}& + \nabla\cdot\left[\rho\textbf{v}\textbf{v} +(p + \frac{B^2}{2})\mathbb{I} - \textbf{B}\textbf{B} \right] = 0,\\
\frac{\partial\mathcal{E}}{\partial t} & + \nabla\cdot\left[(\frac{\rho v^2}{2} + \frac{\gamma}{\gamma-1}p)\textbf{v} + B^2\textbf{v}_T -(\textbf{v}_T\cdot\textbf{B})\textbf{B}\right] = 0,\\
\label{eq:induction}\frac{\partial\textbf{B}_1}{\partial t} &+ \nabla\cdot\left[\textbf{v}_T\textbf{B} - \textbf{B}\textbf{v}_T\right]+ \nabla\psi = 0, \\
\label{eq:divclean}\frac{\partial\psi}{\partial t} &+ c_h^2\nabla\cdot\textbf{B} = -\frac{c_h^2}{c_p^2}\psi,
\end{align}
with $\mathbb{I}$ the identity tensor.
$\mathcal{E} = \rho v^2/2 + B^2/2 + p/(\gamma+1)$ is the total energy density with $\gamma$ the ratio of specific heats (taken to be 5/3 unless otherwise specified).
In Hall MHD, $\mathbf{v}_T = \mathbf{v} + \mathbf{v}_H$ combines the bulk velocity with the Hall velocity $\mathbf{v}_H = -\bar{\delta_i}\mathbf{J}/\rho$.
$\mathbf{J} = \nabla\times\mathbf{B}$ is the current density, which is calculated at the cell faces each timestep using central differencing and averaging \cite{toth08}.
The normalized ion inertial length $\bar{\delta}_i = \delta_i / L_0$ is a set parameter which controls the relative scale of Hall physics to the overall length scale.
In the ideal MHD limit, $\bar{\delta}_i = 0$ and $\mathbf{v}_T \rightarrow\mathbf{v}$.

In equation \ref{eq:divclean}, $c_h$ and $c_p$ are parameters for the propagation and dissipation of local divergence errors.
By default, we set $c_h$ to 95\% of the maximum local, per-cell wavespeed and $c_p$ such that $c_p^2/c_h = 0.18$ (as recommended by \cite{dedner02}).

For numerical stability via the Courant condition, we calculate the global $v_{max}$ using the fast magnetosonic ($c_f$) and whistler wave ($c_w$) speeds \cite{toth08, huba03, marchand18}:
\begin{align}
c_f &= \left[0.5\left(v_A^2 + v_s^2 + \sqrt{(v_A^2+v_s^2)-4v_Av_s}\right)\right]^{1/2}\\
c_w &= \bar{\delta}_i\frac{\pi\|B\|}{2\rho\Delta x} + \sqrt{\left(\bar{\delta}_i\frac{\pi\|B\|}{2\rho\Delta x}\right)^2 + v_A^2}
\end{align}
with $v_A^2 = B^2/\rho$ the normalized Alfv\'en speed and $v_s^2 = \gamma P / \rho$ the normalized sound speed.
Again, ideal MHD is taken in the limit $\bar{\delta}_i = 0$ resulting in $c_w = v_A < c_f$.
The global maximum is then calculated as $v_{max} = \max\left(|v_i| + \max(|c_{f,i}|, |c_{w,i}|)\right)$ taken over all cells $i$.

\subsection{CGL MHD}
\label{sec:CGLMHD}
For the CGL equations (originally developed in \cite{chew1956}), we replace the isotropic ion pressure $p$ with the gyrotropic pressure tensor
\begin{align}
\bar{\bar{p}} = p_\parallel \hat{b}\hat{b} + p_\perp(\mathbb{I} - \hat{b}\hat{b})
\end{align}
The anisotropic pressures $p_\parallel$ and $p_\perp$ are parallel to and perpendicular to the magnetic field unit vector $\hat{b}$, respectively.
This modifies the total energy density to $\mathcal{E} = \rho v^2/2 + B^2/2 + p_\parallel/2 + p_\perp$.

\begin{figure}[t]
    \centering
    \includegraphics[width=0.9\textwidth]{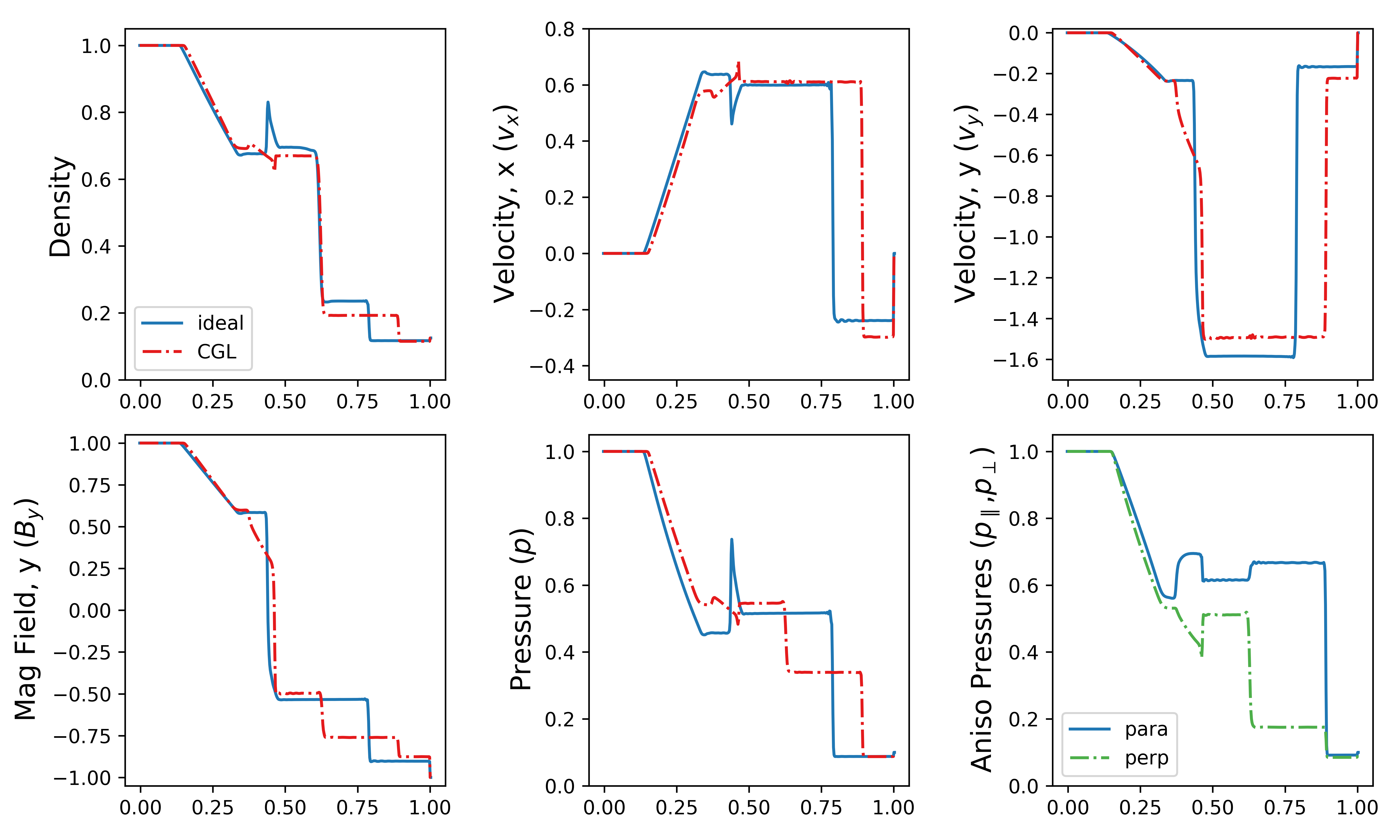}
    \caption{Brio-Wu shock tube at time $t=0.2$ for ideal (solid blue) and CGL (dash-dot red) MHD.
    The bottom-right figure shows the parallel (solid blue) and perpendicular (dash-dot green) pressures for CGL MHD only.
    For CGL MHD, AGATE is able to reproduce the variations in the contact discontinuity region and the selective enhancement of the parallel pressure across the slow shock \cite{luo23,hirabayashi16}.\label{fig:BrioWu}}
\end{figure}

Although one can solve directly for the evolution of the pressure terms (e.g. \cite{meng12}), we use another conserved thermodynamic variable derived from the CGL invariants \cite{chew1956}:
\begin{align}
\frac{d}{dt}\left(\frac{p_\parallel B^2}{\rho^3}\right) = 0~;~\frac{d}{dt}\left(\frac{p_\perp}{\rho B}\right) = 0,
\end{align}
with the convective derivative $d/dt = \partial/\partial t + \mathbf{v}\cdot$.
These two CGL invariants can be combined into a single state variable, $\mathcal{A}$:
\begin{align}
\mathcal{A} = (\rho/B)^3 (p_\perp/p_\parallel)
\end{align}
with the resulting evolution
\begin{align}
\label{eq:Aevo}
\frac{\partial \mathcal{A}}{\partial t}& + \nabla\cdot(\mathcal{A} \textbf{v}) = 0
\end{align}
Previous studies, however, report that this version of $\mathcal{A}$ can be numerically unstable for certain problems.
To prevent negative pressures, a logarithmic formulation can be used instead \cite{santos14, squire23}:
\begin{align}
\mathcal{A}_{\mathrm{log}} = \rho\log\left[\left(\rho^2/B^3\right)~\left(p_\perp/p_\parallel\right)\right]
\end{align}
with the same form of evolution (equation \ref{eq:Aevo}).
AGATE supports both the log and non-log formulations of $\mathcal{A}$, though the log formulation is preferred.

The remaining set of CGL equations are:
\begin{align}
\frac{\partial\rho}{\partial t}& + \nabla\cdot(\rho \textbf{v}) = 0,\\
\frac{\partial\rho\textbf{v}}{\partial t}& + \nabla\cdot\left[\rho\mathbf{v}\mathbf{v} + (p_\perp+B^2/2)\mathbb{I} - \Theta\mathbf{B}\mathbf{B}\right] = 0,\\
\frac{\partial\mathcal{E}}{\partial t} & + \nabla\cdot\left[(\frac{\rho v^2}{2} + \frac{p_\parallel}{2} + 2p_\perp + B^2)\mathbf{v} - \Theta(\mathbf{v}\cdot\mathbf{B})\mathbf{B}\right] = 0,\\
\frac{\partial\textbf{B}_1}{\partial t} &+ \nabla\cdot\left[\textbf{v}\textbf{B} - \textbf{B}\textbf{v}\right]+ \nabla\psi = 0, \\
\frac{\partial\psi}{\partial t} &+ c_h^2\nabla\cdot\textbf{B} = -\frac{c_h^2}{c_p^2}\psi,
\end{align}
with the pressure anisotropy parameter $\Theta \equiv 1 - (p_\parallel - p_\perp)/B^2$ \cite{squire23}.
Note that in the limit of zero magnetic field, the pressure anisotropy vanishes and $\Theta\rightarrow 1$.

Since the numerical algorithm requires both the conserved quantities ($\mathcal{E}$, $\mathcal{A}$) and the primitive quantities ($p_\parallel,~p_\perp$), we present here the conversion between the two forms:
\begin{align}
p_\parallel &= \frac{2\rho^3 e_{\mathrm{int}}}{\rho^3 + 2\mathcal{A}B^3}\\
p_\perp &= \frac{2\mathcal{A}B^3 e_{\mathrm{int}}}{\rho^3 + 2\mathcal{A}B^3}
\end{align}
with the internal energy $e_{int} = \mathcal{E} - \rho v^2 / 2 - B^2/2$.
Similarly, in the log-A case:
\begin{align}
p_\parallel &= \frac{2\rho^2\exp\left[-\mathcal{A_\mathrm{log}}/\rho\right] e_{\mathrm{int}}}{\rho^2\exp\left[-\mathcal{A_\mathrm{log}}/\rho\right] + 2B^3}\\
p_\perp &= \frac{2B^3 e_{\mathrm{int}}}{\rho^2\exp\left[-\mathcal{A_\mathrm{log}}/\rho\right] + 2B^3}
\end{align}

\begin{figure}[t]
    \centering
    \includegraphics[width=0.9\textwidth]{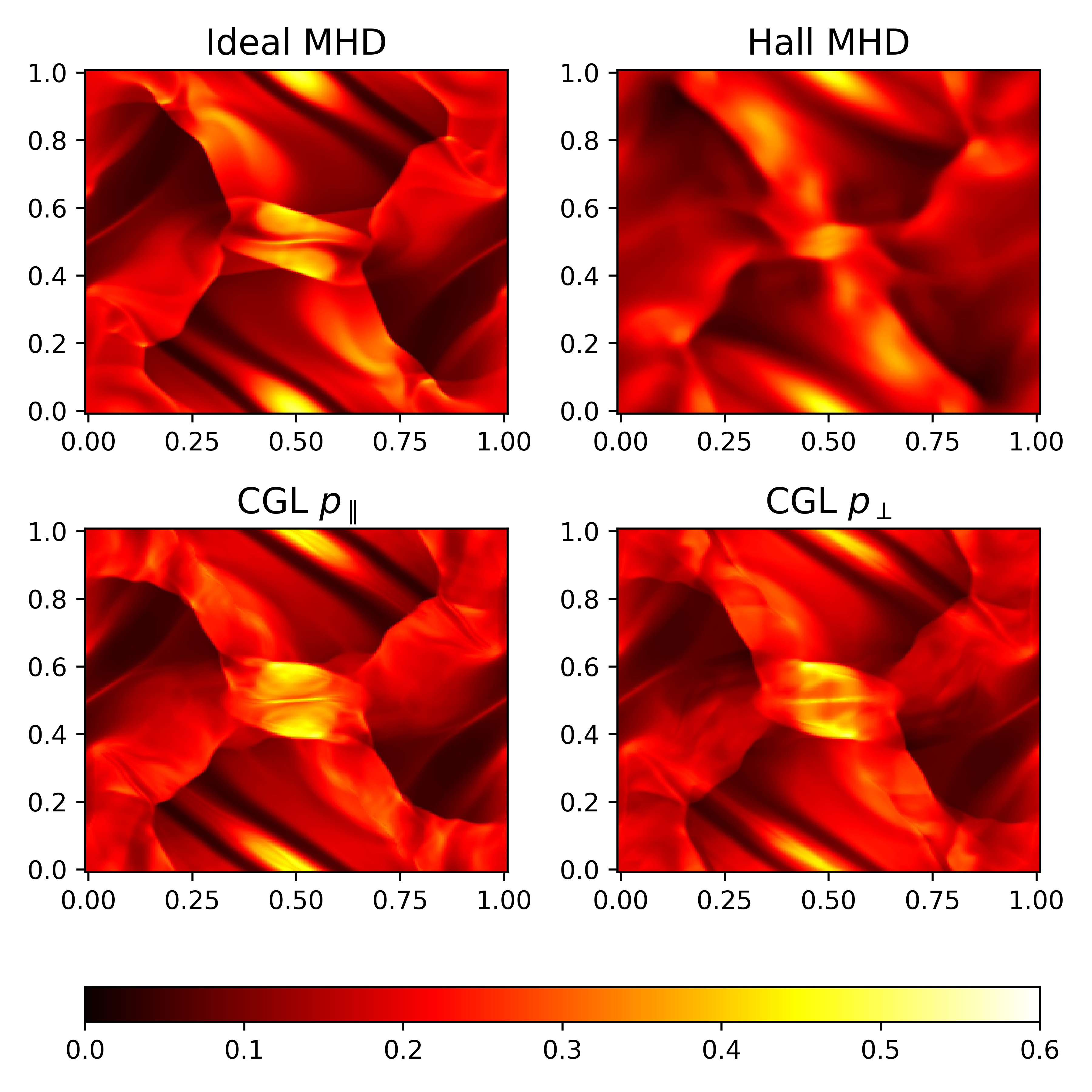}
    \caption{Orszag-Tang Vortex for ideal, Hall, and CGL MHD. Figures shown are of isotropic pressure (top row; ideal: left; Hall: right) and CGL anisotropic pressure (bottom row: $p_\parallel$: left; $p_\perp$: right). Snapshots are taken at $t=0.48$.}\label{fig:OT}
\end{figure}

For the Courant condition, we take the maximum wavespeed
\begin{align}
v_{max} = \max\left(|v_i| + \max\left[|c_{cgl,x}|, |c_{cgl,y}|, |c_{cgl,z}|\right]_i \right),
\end{align}
with the CGL magnetosonic speeds \cite{meng12, hunana19, squire23, santos14, luo23}:
\begin{align}
c_{cgl,d}^2 &= \frac{b_1 \pm \sqrt{b_2^2 + c}}{2\rho}\label{eq:cglFMS}\\
b_1 &= B^2 + 2p_\perp + \cos^2\theta_d\left[2p_\parallel - p_\perp\right]\nonumber\\
b_2 &= b_1 - 6\cos^2\theta_d p_\parallel \nonumber\\
c &= 4 p_\perp^2 \cos^2\theta_d(1-\cos^2\theta_d)\nonumber
\end{align}
with $\cos^2\theta_d = B_d^2 / B^2$ for magnetic component $B_d$ in direction $d=(x,y,z)$.
The fast (slow) magnetosonic speed is taken as the $+$ ($-$) side of eq. \ref{eq:cglFMS}.
Our implementation is closest to the formulation presented in \cite{hunana19}, though all options presented in the citations are equivalent.

Finally, we must consider physical limits for the pressure anisotropy.
The general instabilities considered here are the mirror, firehose, and ion-cyclotron \cite{gary76proton, gary92mirror, gary98proton}:
\begin{align}
-B^2 &\leq P_\perp - P_\parallel \tag{firehose}\\
P_\perp - P_\parallel &\leq \frac{P_\parallel}{p_\perp}\frac{B^2}{2}\tag{mirror}\\
P_\perp - P_\parallel &\leq C_1 P_\parallel\left[\frac{B^2}{2p_\parallel}\right]^{C_2}\tag{ion cyclotron}
\end{align}
For the ion cyclotron instability, $C_1$ and $C_2$ are constants that are generally derived from data \cite{gary94ion, denton94}.
We follow \cite{meng12} and use default values of $C_1=0.3$ and $C_2=0.5$.

Several observational \cite{kasper02,hellinger06,bale09magnetic} and computational \cite{squire23, arzamasskiy23, squire19} studies have shown that plasmas may self-regulate via dynamical pressure-anisotropy feedback, meaning that the anisotropy generally does not exceed the above instability limits.
In CGL MHD, we can imitate this with a point-implicit relaxation condition over the timestep $\Delta t$ \cite{meng12, santos14, luo23}:
\begin{align}
\frac{\delta (\Delta P)}{\delta t} = \frac{\Delta P^\star - \Delta P}{\tau}\label{eq:Preg}
\end{align}
with $\Delta P^\star$ specially chosen to satisfy the above inequalities:
\begin{align}
\Delta P^\star = \max\left[-B^2, \min\left(\Delta P, \frac{P_\parallel}{p_\perp}\frac{B^2}{2}, C_1 P_\parallel\left[\frac{B^2}{2p_\parallel}\right]^{C_2}\right)\right]
\end{align}
We then modify the parallel and perpendicular pressure assuming that the total pressure $P = (p_\parallel + 2p_\perp)/3$ is constant \cite{squire23}:
\begin{align}
p_\parallel^{n+1} = p_\parallel^n  - \frac{2}{3}(\Delta P^\star - \Delta P)\frac{\Delta t}{\Delta t+\tau} \\
p_\perp^{n+1} = p_\perp^n  + \frac{1}{3}(\Delta P^\star - \Delta P)\frac{\Delta t}{\Delta t+\tau}
\end{align}
with the superscripts $n$ and $n+1$ indicating the pressure before and after regularization.
If $\Delta P^\star = \Delta P$, no regularization occurs and the pressures do not change.
$\tau$ is the anisotropy relaxation timescale, which depends on the growth rate of the instabilities for a given plasma state.
By default, we set $\tau$ smaller than the dynamical timescale of the system, i.e. $\tau \approx 10^{-2}\delta t$ or $\Delta t/ (\Delta t + \tau) = 0.99$.
Setting $\tau = 0$ has the effect of applying a "hard" boundary for the pressure anisotropy, as $\Delta P$ will then never be allowed to remain in the unstable region.

\section{Benchmarks}
\label{sec:benchs}
We validate the MHD implementations in AGATE with a series of benchmarks for ideal, Hall, and CGL physics.
The selected benchmarks evaluate both linear and nonlinear behaviors, testing scenarios with smooth solutions and strong shock discontinuities.

\begin{figure}[t]
    \centering
    \includegraphics[width=0.9\textwidth]{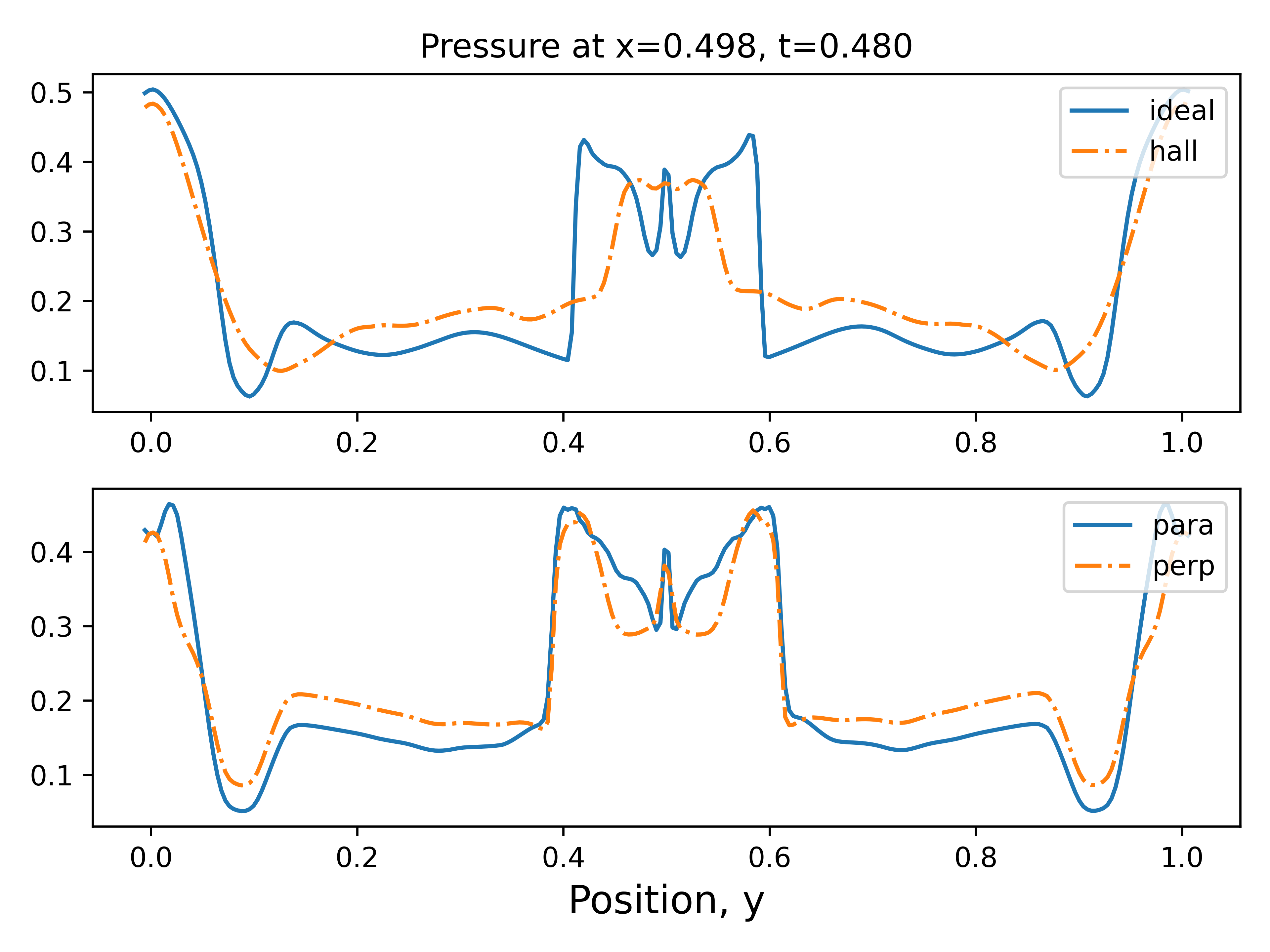}
    \caption{Line profiles of pressure taken from simulations in Figure \ref{fig:OT}. Top plot shows results from ideal (solid blue) and Hall (dot-dashed orange); bottom plot shows CGL $p_\parallel$ (solid blue) and $p_\perp$ (dot-dashed orange). All cuts are taken along $x=0.498$ at $t=0.48$.}\label{fig:OT_cut}
\end{figure}

\subsection{Brio-Wu Shock Tube}
We use the standard Brio-Wu shock tube benchmark for both ideal \cite{briowu} and CGL \cite{hirabayashi16, luo23} MHD to validate handling of non-linear behavior.
We set the domain as $x\in[0,1]$ using $N_x=512$ grid cells, with separate initial conditions for the left ($L$) and right ($R$) states: $\rho_L = 1.0$, $B_{y,L} = 1.0$, $P_L=1.0$; $\rho_R = 0.125$, $B_{y,L} = -1.0$, $P_L=0.1$. $B_x = 0.75$.
All other variables are initialized to zero.
For ideal MHD, we set $\gamma=2.0$. The CGL version of the shock tube is set identically, with zero pressure anisotropy: $p_{\parallel, L/R} = p_{\perp, L/R} = p_{L/R}$.

Figure \ref{fig:BrioWu} demonstrates that AGATE is able to reproduce the usual ideal MHD shock tube behavior as well as the additional changes within CGL MHD\cite{hirabayashi16, luo23}.

\subsection{Orszag-Tang Vortex}
\label{sec:OT}
We use the Orszag-Tang vortex \cite{orszag79, dai98, luo23} to test handling of smooth regions and nonlinear shocks. On a domain of $(x,y)\in[0,1]$ with $N_x = N_y = 256$ grid cells and periodic boundary conditions, the initial conditions are: $\rho = 25/(36\pi)$, $p=5/(12\pi)$, $v_x = -\sin(2\pi y)$, $v_y = \sin(2\pi x)$, $B_x = -B_0\sin(2\pi y)$, and $B_y = B_0\sin(4\pi x)$ with $B_0 = 1/\sqrt{4\pi}$. In the CGL MHD version, we set the initial anisotropy to zero ($p_\parallel = p_\perp = 5/(12\pi)$). For Hall MHD, we additionally set $\delta_i = 10/512$ for 10 cells per ion inertial length.

The results are shown in Figures \ref{fig:OT} and \ref{fig:OT_cut}. AGATE is able to handle complex shocks in all three cases. The Hall MHD run (top right of Figure \ref{fig:OT}) correctly shows smearing of the solution caused by whistler waves.
The CGL MHD run (bottom row) matches results from \cite{luo23}.
Finally, we verify that enforcing zero pressure anisotropy within CGL MHD reduces to the ideal MHD results (not shown here).

\subsection{Whistler wave}
\begin{table}[t]
    \centering
    \begin{tabular}{c c c}
      n & 1D test & 2D test \\ 
      \hline 16 & 0.336 & 0.29 \\
      32 & 0.088 & 0.076 \\
      64 & 0.021 & 0.0176 \\
      128 & 0.0049 & 0.0041 \\
      256 & 0.0012 & 0.00099 \\
      \hline
\end{tabular}
\caption{Relative errors for $B_z$ in the 1D and 2D Whistler wave benchmarks, as calculated by eq. \ref{eq:whiserr}. This demonstrates that the AGATE Hall MHD algorithm is second-order accurate.}\label{tab:whisterrs}
\end{table}
To further verify the Hall MHD algorithm, we initialize 1D and 2D whistler waves \cite{toth08, bard16ph} with right-hand circular polarization.
The grid is $x\in[-100,100]$ with periodic boundary conditions.
Among a constant background of $\rho = 1.0$, $B_x=100$, $p=1.0$, we define a wave $v_y = -\delta_v\cos(k_x x)$, $v_z = \delta_v\sin(k_x x)$, $B_y = \delta_b\cos(k_x x)$, $B_z = -\delta_b\sin(k_x x)$ with $\delta_b = 0.001$ and $\delta_v = \delta_b B_x/(v_\phi\rho)$.
The wavelength is set to 200 such that $k_x = 2\pi/200$.
The (normalized) phase speed of the whistler wave is
\begin{align}
c_\phi = \frac{\delta_i v_A k}{2} + \sqrt{v_A^2+\frac{\delta_i^2 v_A^2 k^2}{4}}
\end{align}
with the normalized Alfv\'en speed $v_A = B_x/\sqrt{\rho} = 100$ and we set $\delta_i = 35.1076$ such that $v_\phi = 169.345$ and the whistler period $\tau_W = 200/v_\phi \approx 1.18102$.

To check the accuracy of AGATE, we calculate the numerical error via:
\begin{align}
E_n = \frac{\Sigma_{i=1}^n|B_{z,i} (\tau_W)- B_{z,i}(0)|}{\Sigma_{i=1}^n |B_{z,i}(0)|}\label{eq:whiserr}
\end{align}
There is no special reason for using $B_z$; any transverse component of velocity or magnetic field will produce similar results \cite{toth08}.
As shown in Table \ref{tab:whisterrs}, the relationship between error and numerical resolution demonstrates that AGATE (under the current default settings) is second-order accurate.

An extension to this test is to rotate the whistler wave by an angle $\alpha$ and simulate in 2D.
Following \cite{toth08}, we select $\alpha = \tan^{-1}0.5 = 26.56^\circ$ in order to maintain a (1,2) translational symmetry.
As a consequence, the initial velocity changes to $v_x = \delta_v\cos(k_xx+k_yy)\sin(\alpha)$, $v_y = -\delta_v\cos(k_xx+k_yy)\cos(\alpha)$, $v_z = \delta_v\sin(k_xx+k_yy)$.
The initial magnetic field rotates both the background and the perturbed field: $B_x = B_{x,0}\cos{\alpha}-\delta_b\cos(k_xx+k_yy)\sin{\alpha}$, $B_y = B_{x,0}\sin{\alpha}+\delta_b\cos(k_xx+k_yy)\cos{\alpha}$, $B_z = -\delta_b\sin(k_xx+k_yy)$, with $B_{x,0} = 100$ the same as for the 1D problem.
To maintain periodicity in the $x$-direction, the grid must be lengthened to $x,y\in[-100/\cos\alpha, 100/\cos\alpha]$.
Additionally, we apply a sheared zero-gradient boundary in $y$: the ghost cell at index $(i,j)$ is set equal to the ghost cell at $(i\mp1, j\pm2)$ for the boundary at the down-side (up-side) of the simulation, wrapping around periodically in the $x$ direction as necessary.

We repeat the error calculation (eq. \ref{eq:whiserr}) for $B_z$; the results (Table) demonstrate second-order accuracy in the Hall MHD algorithm.

\subsection{CGL Magnetosonic Waves}
\begin{figure}[t]
    \centering
    \includegraphics[width=0.9\textwidth]{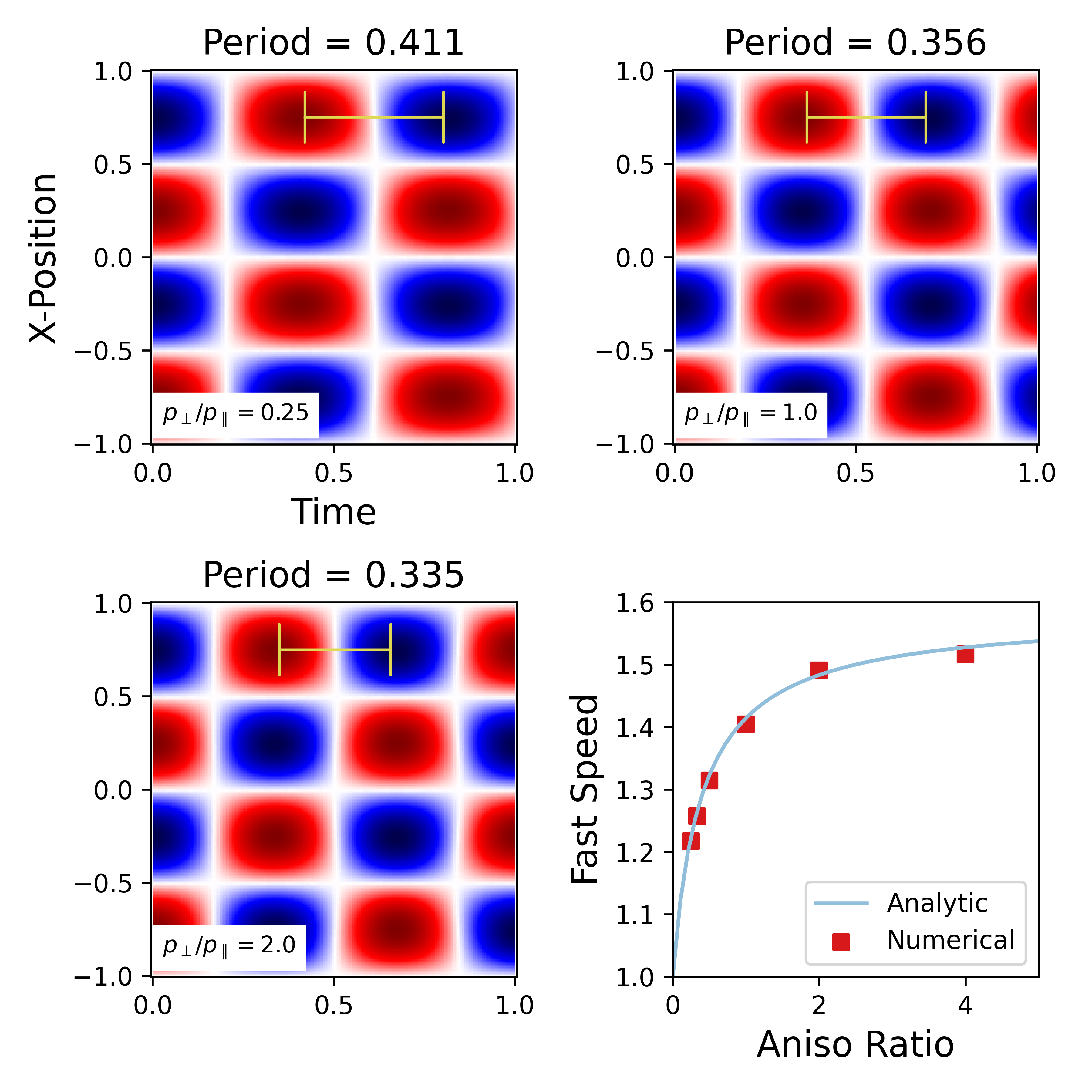}
    \caption{The first three panels show keograms of $v_x$ as a function of position $x$ (horizontal) and time $t$ (vertical) for the CGL standing magnetosonic wave test for three selected anisotropy ratios. Color shows magnitude of $v_x$ (blue: negative, red: positive). The lower right panel shows a comparison between the theoretical fast magnetosonic wavespeed (see text) and the numerical value derived from observed periods.
    \label{fig:CGLMstand}}
\end{figure}
We test the CGL implementation via a standing \cite{luo23} and a propagating \cite{meng12} magnetosonic wave.

First, the standing wave is initialized on a domain $x\in[-1,1]$ with $N_x = 256$ grid cells with periodic boundary conditions.
The background state is $\rho=1$, $B_y=1$, and $P = 0.5$ with a perturbation $v_x = 0.01\sin(2\pi x)$.
The wave is parameterized via the anisotropy ratio $a_p = p_\perp/p_\parallel$.
Using $P = (p_\parallel+2p_\perp)/3$, we obtain the parallel and perpendicular pressures directly from $a_p$: $p_\parallel = 3P / (1 + 2  a_p)$ and $p_\perp = a_p p_\parallel$.
All other variables are set to zero and pressure regularization is turned off for this test.
\begin{figure}[t]
    \centering
    \includegraphics[width=0.9\textwidth]{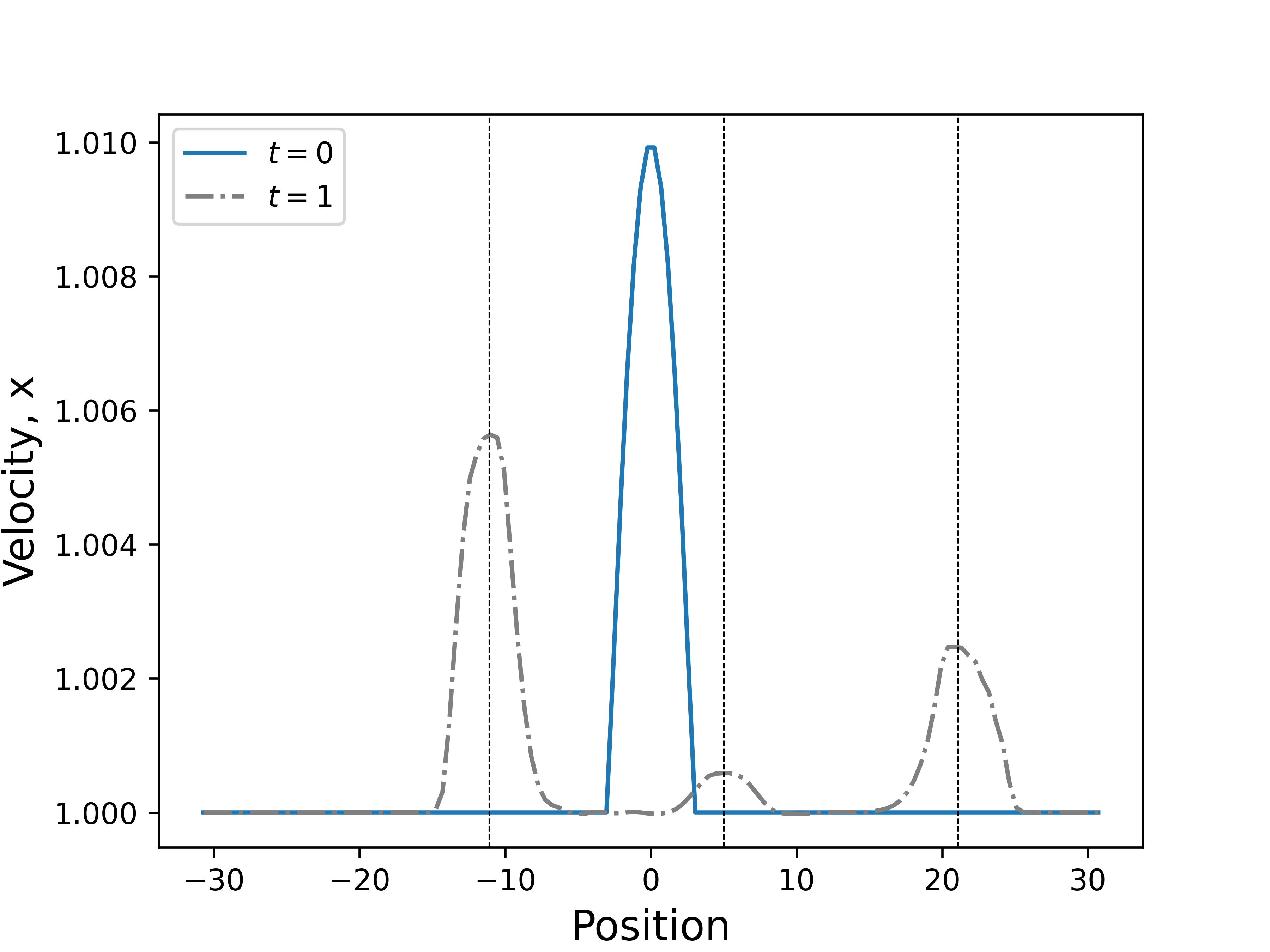}
    \caption{CGL magnetosonic propagation test. The initial perturbation (solid blue) and the evolved state at $t=1$ (gray dot-dashed). The three vertical dotted black lines, from left to right, correspond to the theoretical distances ($d = vt = v$) at $t=1$ for the left-going fast mode ($v_x - v_{ms}$), the slow mode ($v_x$), and the right-going fast mode ($v_x + v_{ms}$).\label{fig:CGLMprop}}
\end{figure}

The resulting magnetosonic wavespeed can be calculated from eq. \ref{eq:cglFMS}, noting that $\cos^2\theta_x = B^2_x/B^2=0$: $c_{ms} = \sqrt{B^2/\rho~+~2p_\perp/\rho}$.
For a set of anisotropy ratios, we calculate the simulated period of the wave and compare with its theoretical calculation: $\tau_{ms} = \lambda/c_{ms}$ with the wavelength $\lambda=1$.
Following \cite{luo23}, we illustrate the magnetosonic wave with keograms showing $v_x$ as a function of time and position.
The comparison of the theoretical and numerical phase speeds are shown in Figure \ref{fig:CGLMstand}.

The propagating magnetosonic wave is initialized on a grid $x\in[-30,30]$ with a background state $\rho=1$, $v_x=5$, $B_y=15$, $p_\parallel = 14$, and $p_\perp = 17$ and all other variables set to zero.
Next, a localized perturbation of the density, magnetic field, and pressure is created between $-3<x<3$: $\delta\rho = 0.01\cos(2\pi x / 12)$, $\delta{p_\parallel} = 0.14\cos(2\pi x / 12)$, $\delta{p_\perp} = 0.17\cos(2\pi x / 12)$, and $\delta{B_y} = 0.15\cos(2\pi x / 12)$.

This perturbation will generate three waves: two fast magnetosonic waves propagating with speeds $v_x \pm v_{ms} = 5 \pm 16.09 = (21.09, -11.09)$, and one slow magnetosonic wave propagating at speed $v_x=5$.
We verify this in Figure \ref{fig:CGLMprop} showing the positions of the three waves at $t=1$.

\subsection{CGL Firehose Instability}
\begin{figure}[t]
    \centering
    \includegraphics[width=0.9\textwidth]{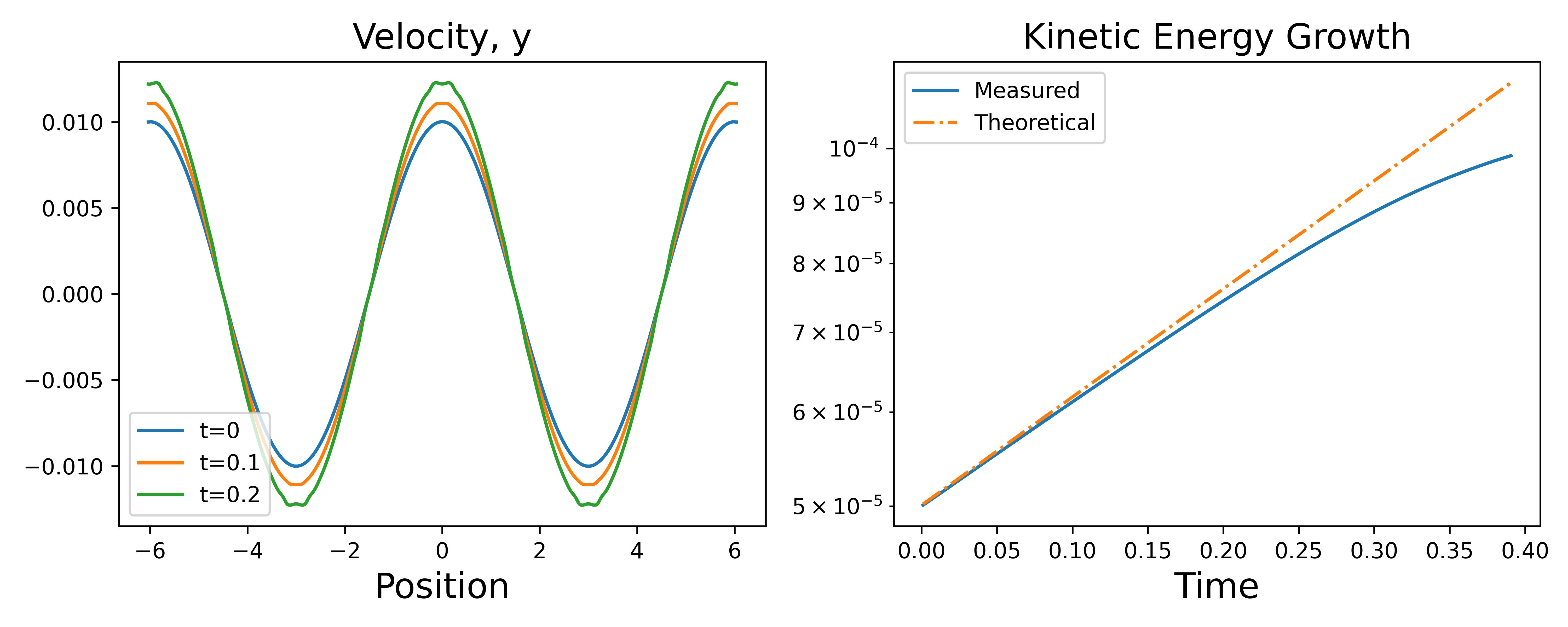}
    \caption{Left: Velocity growth caused by the firehose instability. Small oscillations at the wave extrema can be seen for $t=0.2$; these overtake the growth of the firehose instability at later times. Right: Comparison of theoretical and simulation growth rate of the $y$-kinetic energy, $\rho v_y^2/2$, as a function of time. Note that $v_y^2 \propto \exp[-2i\omega t] = \exp[2(1.0472) t]$ as plotted here. (See text for the calculation of the growth rate $\omega$.)\label{fig:CGLfire}}
\end{figure}
With this test, we check that turning off the pressure regularization reproduces the firehose instability in the appropriate limit.
Following \cite{meng12}, we initialize an Alfv\'en mode on a grid $x\in[-6,6]$ with $N_x = 512$ cells.
The background initial conditions are $\rho=1$, $B_x=10$, $p_\parallel = 104$, $p_\perp = 3$ with perturbed variables $v_y = 0.01\cos(2\pi x/6)$ and $B_y = 0.1\cos(2\pi x/6 + \pi/2)$.
All other variables are set to zero, and we enforce no pressure regularization by setting $\delta t/(\delta t+\tau) = 0$ (effectively setting $\tau\rightarrow\infty$ in eq. \ref{eq:Preg}).

As a result, the pressure anisotropy exceeds the firehose threshold and has a (normalized) growth rate $\omega = k_\parallel v_A\sqrt{1+(p_\perp-p_\parallel)/B^2} \approx 1.0472i$ such that $v_y \propto \exp[-i\omega t] = \exp[1.0472t]$.
We confirm that AGATE is able to reproduce this behavior (Figure \ref{fig:CGLfire}).
We note that, like \cite{meng12}, we observe small oscillations near the extrema caused by numerical errors.
These cause short wavelength errors which eventually cause deviations from the firehose instability.

\subsection{GEM Reconnection}
\label{sec:GEMbench}
The GEM reconnection challenge \cite{GEM2001} provides a well-established benchmark for ideal, Hall, and anisotropic \cite{birn01b} MHD.
The simulation begins with a unpertubed Harris sheet equilibrium \cite{harris62} on a computational domain with dimensions $L_x = 25.6$ and $L_y=12.8$, centered at the origin, using $N_x = 512$ and $N_y = 256$.
The initial equilibrium is $B_x = \tanh{2y}$, $\rho = 0.2 + \sech^2(2y)$, $p = 0.5\rho$ with all other variables initialized to zero.
A magnetic perturbation is then introduced via:
\begin{align}
\delta B_x = -\delta_B \left(\pi/L_y\right) \cos(2\pi x / L_x)\sin(\pi y / L_y) \nonumber\\
\delta B_y = +\delta_B \left(2\pi/L_x\right) \sin(2\pi x / L_x)\cos(\pi y / L_y)
\end{align}
with $\delta_B = 0.1$.
For the CGL run, we initialize a isotropic pressure: $p_\parallel = p_\perp = p$ and set the regularization factor $\Delta t/(\Delta t + \tau) = 0.99$.

\begin{figure}[t]
    \centering
    \includegraphics[width=0.9\textwidth]{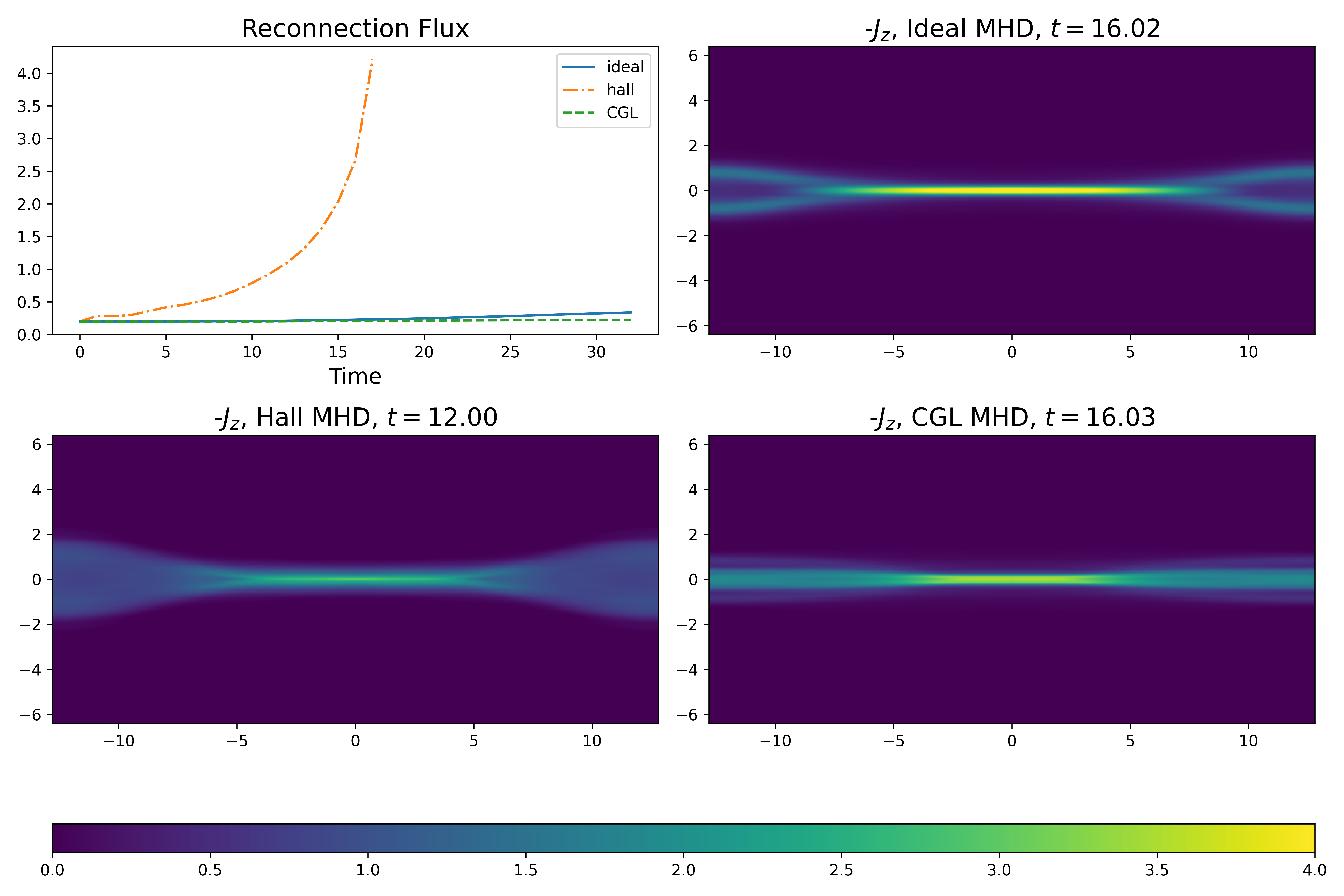}
    \caption{Top left: GEM reconnection flux as a function of time (c.f. \cite{GEM2001}).
    Others: Negative of Current Density $z$-component for ideal (top right), Hall (bottom left), and CGL (bottom right) MHD at selected times.
    \label{fig:CGLGEM}}
\end{figure}

Figure \ref{fig:CGLGEM} shows the simulation results.
The reconnected flux, calculated as $-\int_{-L_x/2}^0 B_y dx$, is shown as a function of time for each variant of MHD (top left subfigure).
The results align with previous findings \cite{GEM2001}, demonstrating enhanced reconnection rates when including the Hall term, while ideal and CGL MHD exhibit similar, slower rates.
The remaining panels display the current sheet structure for all three cases.
While our results are consistent with previous studies \cite{birn01b,luo23}, our current sheet appears thicker than that reported by \cite{luo23}.
We anticipate that this is likely due to higher numerical resistivity in our second-order scheme compared to the higher-order reconstruction methods used in GAMERA \cite{zhang2019gamera}.

\subsection{Timings}
\label{timing}
We conducted performance benchmarks to evaluate AGATE's various acceleration options, noting that our primary focus is to use the GPU options for high-performance computing applications.
Using the Orszag-Tang vortex test case (Section \ref{sec:OT}), we simulated ideal MHD with divergence cleaning to time $t=0.5$ across grid resolutions ranging from $32^2$ to $2048^2$.
All simulations used default algorithm settings and were executed on a system featuring an AMD EPYC CPU (2.654 MHz; 2 sockets; 64 cores/socket; 2 threads/core; 256 threads total) and an NVIDIA A100 GPU.
For the \texttt{Numba} acceleration option, preliminary testing on a $256^2$ grid identified optimal performance with our system CPU using 16 threads.
Comparative timing results are presented in Table \ref{tab:timings}.

\begin{table}[t]
    \centering
    \begin{tabular}{c c c c c c c c c c}
      n & \texttt{None} & \texttt{Numpy} & \texttt{Numba} & \texttt{Cupy} & \texttt{Kernel} & \texttt{KernX3} & \texttt{Ath++} & \texttt{AthOMP} & \texttt{ChollaX3}\\ 
      \hline
      $32^2$ & 28.8 & 0.28 & 13.3 & 0.73 & 0.27 & 0.33 & 0.048 & 0.065& 0.76\\
      $64^2$ & 231 & 1.46 & 13.6 & 1.37 & 0.450 & 0.49& 0.362& 0.204&0.84\\
      $128^2$ & 1802 & 17.6 & 16.4 & 2.63 & 0.82 & 0.98& 2.74& 0.985&1.17\\
      $256^2$ & N/A & 122 & 49.5 & 5.17 & 1.55& 2.66 & 22.1& 4.26&3.16\\
      $512^2$ & N/A & 895 & 247 & 10.3 & 3.04 & 19.0& 177.0& 24.4&18.0\\
      $1024^2$ & N/A & 7477 & 1320 & 20.3 & 12.0 & 147& 1396& 214&132\\
      $2048^2$ & N/A & N/A & N/A & 93.4 & 87.1 & 1172 & N/A& 1003& 1021\\
      \hline
    \end{tabular}
    \caption{Performance comparison of AGATE acceleration modes for ideal MHD Orszag-Tang vortex simulations (with divergence cleaning) to $t = 0.5$.
    Timings are reported in seconds. CPU-based modes (\texttt{None}, \texttt{Numpy}, \texttt{Numba}) were executed on an AMD EPYC CPU with \texttt{Numba} utilizing 16 threads.
    GPU-based modes (\texttt{Cupy}, \texttt{Kernel}) employed an NVIDIA A100 GPU.
    \texttt{KernX3} is performed on a $N^2x3$ grid in order to match \texttt{ChollaX3}, which can only operate with a 3D grid.
    \texttt{Ath++} is \texttt{Athena++} run on a single node, while \texttt{AthOMP} utilizes OpenMP with parameter \texttt{num\_threads} set to 16.
    Note: \texttt{Numba} requires a one-time 13-second compilation overhead for each run.
    "N/A" indicates that the benchmark was not run due to prohibitively long execution times.}\label{tab:timings}
\end{table}

Our performance testing reveals that while a pure Python implementation is impractical for high-performance computing, using acceleration libraries result in substantial improvements.
\texttt{Numpy} alone provides about a 100-fold speedup over pure Python for a $128^2$ grid.
\texttt{Numba}, despite requiring a 13-second initial JIT compilation overhead, achieves 2-5 times better performance than \texttt{Numpy} for sufficiently large problems.
The GPU-based options demonstrate the highest performance, with speedups of $>$40 times over \texttt{Numba} and $>$300 times over \texttt{Numpy} for the larger benchmarks.

We compare our timing results with established HPC codes in the literature: the CPU-based Athena++ \cite{Stone2020,athena24} and the GPU-based Cholla \cite{schneider15,caddy24}. 
Using the same Orszag-Tang benchmark across identical problem sizes, we followed the implementation instructions provided in each code's repository. 
Athena++ supports a OpenMP implementation; for this mode, we set the size of the individual MeshBlocks such that the parameter \texttt{num\_threads} was equal to 16 for all problem sizes.
Since Cholla requires a grid with at least three dimensions, we used an NxNx3 grid for the benchmark. 
For a more equitable comparison with AGATE, we also executed our Kernel benchmark with a similar 3D grid configuration (\texttt{KernelX3}).

An advantage of our Python-based approach over the C++ implementations of Athena++ and Cholla is the virtual elimination of compilation time. 
AGATE's only preprocessing overhead resulted from JIT compilation, which took approximately 13 seconds in our testing environment for the Numba option.
AGATE's GPU options had minimal ($<<$ 1 second) overheads, and the other CPU options had zero overhead.
This compares favorably to Athena++ and Cholla, which required 81 and 119 seconds to compile, respectively.

Our results clearly show that Athena++ is considerably more optimized for CPU-based HPC operations than AGATE, which is expected given our focus on GPU implementations. 
Nevertheless, AGATE's minimal compilation requirements and seamless integration with Python workflows make it particularly suitable for rapid deployment, especially for students in learning environments or when prototyping with smaller problem sizes.

For GPU performance, we found that Cholla executes approximately 10-15\% faster than AGATE at the largest problem sizes, while performance is comparable at smaller scales. 
We consider this modest speed difference an acceptable trade-off for the advantages of a Python-based implementation. 
Overall, our results demonstrate that AGATE effectively supports high-performance computing applications when leveraging GPU acceleration.

\section{Conclusion}
In this work, we introduce AGATE, a Python-based framework designed for solving equations using a modular, object-oriented approach.
AGATE employs an object-oriented architecture that separates interface specifications from numerical implementations, enabling flexible adaptation to different numerical methods.
This design philosophy allows users to modify or extend individual components without disrupting the overall solver architecture.

By default, the framework implements a Godunov-type finite-volume scheme for solving the magnetohydrodynamic (MHD) equations with support for the ideal, Hall, and CGL variants.
The numerical algorithm operates through modular classes that handle reconstruction, wave speed calculations, flux evaluation, and source terms.
AGATE provides utility classes to streamline common simulation tasks from problem initialization to runtime analysis, while offering automated algorithm setup functions for user convenience.
This balance between automation and customization makes AGATE suitable for multiple use cases: rapid prototyping, large-scale HPC simulations, numerical algorithm development, and plasma physics education.

We validate AGATE's MHD implementations through a comprehensive suite of benchmarks covering the ideal, Hall, and CGL MHD equations.
These benchmarks assessed both linear and nonlinear numerical performance, encompassing scenarios with smooth solutions and cases involving strong shock discontinuities.
The results demonstrate that AGATE accurately captures essential physical phenomena across different MHD regimes.

Performance testing reveals that while pure Python (acceleration option: \texttt{None}) is impractical for most uses, the \texttt{Numpy} and \texttt{Numba} CPU options provide viable speedups for small simulations.
GPU acceleration through \texttt{Cupy} and \texttt{Kernel} yields further performance gains, achieving speedups of $>$40x over Numba for large problem sizes.
These results demonstrate that Python-based scientific computing can achieve competitive performance when properly optimized.

Looking forward, AGATE's modular architecture provides a foundation for future extensions, including additional physics modules, more sophisticated numerical methods, and support for advanced grid geometries.
The demonstrated performance and extensibility establish AGATE as a viable platform for high-performance MHD simulations in research, application, and educational contexts.

\section*{Acknowledgments}
This project was funded by the NASA Living with a Star Strategic Capabilities (LWSSC) program.
Software packages used in this paper include: Matplotlib \cite{matplotlib07}; Numpy \cite{numpy20}; CuPy \cite{cupy17}, Numba \cite{numba15}.
The AGATE open-source repository can be found at: \url{https://git.smce.nasa.gov/marble/agate-open-source}

\appendix
\section{Code Differences between AGATE Acceleration Options}
\label{app:acclr}
As mentioned in Section \ref{sec:acclr}, there are five main options for running AGATE: \texttt{None}, \texttt{Numpy}, \texttt{Numba}, \texttt{Cupy}, and \texttt{Kernel}.
Each AGATE Implementation class (Section \ref{sec:modular}) defines several functions, each compatiable with one (or more) of these options.
When the acceleration option is passed in, AGATE automatically selects the appropriate function as part of its start-up procedure.

We now give code examples to illustrate differences between options.
(Although the examples here are for 1D arrays, the principles are generalizable for multi-dimensional arrays.)
First, using \texttt{None} implements the calculation loop in pure Python, running on a CPU:

\begin{lstlisting}[language=python]
def example(nx, array1, array2, array3):
    # 'prange' instead of 'range' for compatibility with Numba
    for i in prange(nx):
        intermediate = math.sin(array2[i] - 0.1)
        if intermediate < 0:
            intermediate = -1.
        array3[i] = 0.5*(array1[i]+array2[i]) * intermediate
\end{lstlisting}

Note that for maximum compatibility with Numba, we use \texttt{numba.prange} instead of Python \texttt{range}.
\texttt{prange} has the same effect as \texttt{range} when run in Python-only mode.

In the \texttt{Numba} option, this basic loop is fed to Numba's just-in-time (JIT) compiler for parallelization with individual CPU cores:
\begin{lstlisting}[language=python]
exampleNumba = numba.jit(example, nopython=True, parallel=True)
# function call
exampleNumba(nx, array1, array2, array3)
\end{lstlisting}

The final CPU-only option, \texttt{Numpy}, uses Numpy's built-in vectorization for array operations.
This necessitates writing a different, loop-less function:
\begin{lstlisting}[language=python]
def exampleNumpy(array1, array2, array3):
    intermediate = np.sin(array2 - 0.1)
    intermediate[intermediate < 0.] = -1.
    array3[:] = 0.5*(array1 + array2) * intermediate
\end{lstlisting}
We note that Cupy does have drop-in replacement for accelerating Numpy functions on the GPU.
However, preliminary testing showed that this option was inferior to the other GPU options, so AGATE does not implement it.

Among the GPU options AGATE supports, \texttt{Cupy} employs JIT compilation of a pure Python loop, similar to the CPU-only approach used by Numba. However, a separate loop function must be created to enable GPU thread indexing, as this instruction is incompatible with CPU operation:
\begin{lstlisting}[language=python]
def exampleGPU(nx, array1, array2, array3):
    i = cupyx.jit.threadIdx.x + cupyx.jit.blockIdx.x * cupyx.jit.blockDim.x
    if i < nx:
        intermediate = math.sin(array2[i] - 0.1)
        if intermediate < 0:
            intermediate = -1.
        array3[i] = 0.5*(array1[i]+array2[i]) * intermediate

example_Cupy_Jit = cupyx.jit.rawkernel()(exampleGPU)
# function call
example_Cupy_Jit[blockDim, threadDim](nx, array1, array2, array3)
\end{lstlisting}
Although Numba also has built-in GPU JIT support, preliminary testing found Cupy's JIT consistently produced faster kernels for effectively the same code.
As a result, although AGATE does technically support Numba+GPU, we prefer using Cupy for GPU-accelerated simulations.

Finally, for AGATE's \texttt{Kernel} option, Cupy provides the option to write, compile, and call low-level kernel code:
\begin{lstlisting}[language=python]
instructions = """
extern "C"{
__global__ void kernExample(int nx, double *array1, double *array2, double *array3){
    int i = blockIdx.x*blockDim.x + threadIdx.x;
    double intermediate;

    if (i < nx){
        intermediate = sin(array2[i] - 0.1);
        if(intermediate < 0.){
            intermediate = -1.;
        }
        array3[i] = 0.5 * (array1[i] + array2[i]) * intermediate;
    }
}
}
"""

exampleKernel = cupy.RawKernel(instructions, "kernExample")
# function call
exampleKernel((blockDim,), (threadDim,), (nx, array1, array2, array3))
\end{lstlisting}


\bibliographystyle{elsarticle-num-names}
\bibliography{AGATE_paper.bib}

\end{document}